\documentclass[a4paper,11pt]{article}
\pdfoutput=1 

\usepackage{lineno}
\usepackage{jinstpub} 

\title{Nanodiamond photocathodes for MPGD-based single photon detectors at future EIC}

\author[f]{F.~M.~Brunbauer,}
\author[b]{C.~Chatterjee,}
\author[d]{G.~Cicala,}
\author[e]{A.~Cicuttin,}
\author[b]{P.~Ciliberti,}
\author[e]{M.~L.~Crespo,}
\author[b]{D.~D`Ago,}
\author[a]{S.~Dalla~Torre,}
\author[a]{S.~Dasgupta,}
\author[a]{M.~Gregori,}
\author[c]{T.~Ligonzo,}
\author[a]{S.~Levorato,}
\author[f,g]{M.~Lisowska,}
\author[a]{G.~Menon,}
\author[a]{F.~Tessarotto,}
\author[f]{L.~Ropelewski,}
\author[e,1]{Triloki,\note{Corresponding author.}}
\author[c]{A.~Valentini,}
\author[d]{L.~Velardi,}
\author[a,2]{Y.~X.~Zhao \note{Present address: Institute of Modern Physics, Chinese Academy of Sciences, Lanzhou, 730000, China}}

\affiliation[a]{INFN Trieste,\\ Trieste, Italy}
\affiliation[b]{University of Trieste and INFN Trieste,\\ Trieste, Italy}
\affiliation[c]{University Aldo Moro of Bari and INFN Bari,\\ Bari, Italy}
\affiliation[d]{CNR-ISTP and INFN  Bari,\\ Bari, Italy}
\affiliation[e]{Abdus Salam ICTP, Trieste, Italy and INFN Trieste,\\ Trieste, Italy}
\affiliation[f]{European Organization for Nuclear Research (CERN), CH-1211 Geneve 23, \\Switzerland}
\affiliation[g]{Wrocław University of Science and Technology, Wybrzeże Wyspiańskiego 27, 50-370 Wrocław, \\Poland}

\emailAdd{Triloki@ts.infn.it}

\abstract{We are developing gaseous photon detectors for Cherenkov imaging applications in the experiments at the future Electron Ion Collider. CsI, converting photons in the far ultraviolet range, is, so far, the only photoconverter compatible with the operation of gaseous detectors.   It is very delicate to handle due to its hygroscopic nature: the absorbed water vapour decomposes the CsI molecule.  In addition, its quantum efficiency degrades under  ion bombardment. These are the key reasons to quest for novel, less delicate materials for photocathodes adequate for gaseous photon detectors. Layers of hydrogenated nanodiamond particles have recently been proposed as an alternative material and have shown promising characteristics. The performance of nanodiamond photocathodes coupled to thick GEM based detectors is the objects of our ongoing R\&D. 
\par
The first phase of these studies includes the characterization of thick GEM coated with nanodiamond layers and the robustness of its photoconverting properties with respect to the bombardment   by ions from the multiplication process in the gaseous detector.
\par
The approach is described in detail as well as all the results obtained so far within these exploratory studies.
}


\proceeding{International conference on Instrumentation for Colliding Beam Physics: INSTR20\\
  24 to 28 February, 2020\\
  Cohosted by the Budker Institute of Nuclear Physics (BINP) and Novosibirsk State University (NSU), Novosibirsk, Russia}

\begin{document}
\maketitle
\flushbottom
\section{Introduction}
\label{sec:intro}
The future Electron Ion Collider (EIC)~\cite{EIC} is the facility dedicated to understanding quantum chromodynamics (QCD), including the elusive non-perturbative effects and the answer to key questions, pending since long. Among them: the origin of nucleon mass and spin  and the properties of dense gluon systems. The experimental activity at EIC requires  efficient hadron Particle IDentification (PID) in a wide momentum range, including the challenging scope of hadron PID at high momenta, namely larger than $6$-$8~GeV/c$. A gaseous Ring Imaging CHerenkov (RICH) is the only possible choice for this specific task. The number of Cherenkov photons generated in a light radiator is limited. In spectrometer setups,  these number of photons is recovered by using long radiators. The compact design of the experimental setup at the EIC collider imposes limitations on the radiator length, requiring a dedicated strategy. In the far UltraViolet (UV) spectral region ($\sim120$~nm), the number of generated Cherenkov photon increases, according to the Frank-Tamm distribution~\cite{Frank:1937fk}. This suggests the detection of photons in the very far UV range. The standard fused-silica windows are opaque for wavelengths below 165~nm. Therefore, a windowless RICH is a potential option. The approach also points to the use of gaseous photon detectors operated with the radiator gas itself~\cite{windowless-RICH}.
\par 
The MicroPattern Gaseous Detector (MPGD)-based Photon Detectors (PD) have recently been demonstrated as effective devices~\cite{PM18} for the detection of single photon in  Cherenkov imaging counters. These PDs are composed of a hybrid structure, where two layers of THick GEM (THGEM) multipliers~\cite{thgem} are followed by a MICRO-MEsh GAseous Structure (MICROMEGAS)~\cite{mm} stage; the top layer of the first THGEM is coated with a reflective CsI PhotoCathode (PC).
\par 
CsI PC is, so far, the only feasible option for gaseous detectors, thanks to its relatively high work function that makes it more robust than other ones commonly
used in vacuum-based detectors.  CsI has high Quantum Efficiency (QE) in the far UV spectral region. In spite of its successful application~\cite{RD26},  it presents  problematic aspects.  It is hygroscopic: the absorbed water vapour splits the CsI molecule, causing a degradation in QE~\cite{NIMA_695_2012_279}. Therefore,  the handling of CsI PC is a very delicate operation. QE degradation also appears after intense ion  bombardment, when the integrated charge is  $1~mC/cm^{2}$~\cite{NIMA_574_2007_28} or larger. In gaseous detectors, ion bombardment of the cathode is by the ion avalanche produced in the multiplication process. The fraction of ions reaching the cathode depends on the detector architecture. In recent years, MPGD schemes with enhanced ion blocking capability have been developed~\cite{PM18, ibf-blocking}.
\par
The search for an alternative UV sensitive photocathode overcoming these limitations is therefore an important goal for the research and development (R\&D) program for the experiments at the EIC. In the present article, we  present the preliminary results on  NanoDiamond (ND) and Hydrogenated-NanoDiamond (H-ND) coated THGEM detectors. They also include    preliminary  results about QE robustness with respect to ion bombardment.


\section{Nanodiamond based PC as an alternative of CsI PC}
The high QE value of CsI photocathode makes it the mostly used photoconverter for the UV sensitive devices. This high QE value is related to its low electron affinity ($0.1~eV$) and wide band gap ($6.2~eV$) ~\cite{JAP_77_1995_2138}. The ND particles have a comparable band gap of $5.5~eV$ and low electron affinity of $0.35$-$0.50~eV$. H-NDs exhibit chemical inertness and radiation hardness. ND hydrogenation lowers the electron affinity to -$1.27~eV$. The Negative Electron Affinity (NEA) allows an efficient escape into vacuum of the generated photoelectrons without an energy barrier at the surface~\cite{NDRep-1}. A novel ND hydrogenation procedure,  developed
in Bari~\cite{NDRep-1, NDreport}, provides high and stable QE. A comparison of CsI and ND QE can  be extracted from  literature ~\cite{NDRep-1, NIMA_502_2003_76}.  

\section{The R\&D activity}

\subsection{THGEM characterization}

\par The initial phase of our R\&D studies consisted in  coating five THGEMs with ND and H-ND powder. \par

THGEMs are robust gaseous electron multipliers based on GEM principle scaling the geometrical parameters. 
It is obtained via standard PCB drilling and etching processes. The 35~$\mu$m copper layer is coated with $\approx$5~$\mu$m of Ni, followed by 200~nm Au. The THGEMs used for our studies have an active area of $30\times30~mm^{2}$ with a  hole diameter of  0.4~mm, a pitch of 0.8~mm and a thickness of 470~$\mu$m. THGEMs with different rim i.e. the clearance  ring around  the hole edge have been used: $\le5\mu$m (no rim), $\sim10~\mu$m and $\sim20~\mu$m. 
\par

\begin{figure}
\begin{minipage}[c]{1\textwidth}
    \includegraphics[width=\textwidth]{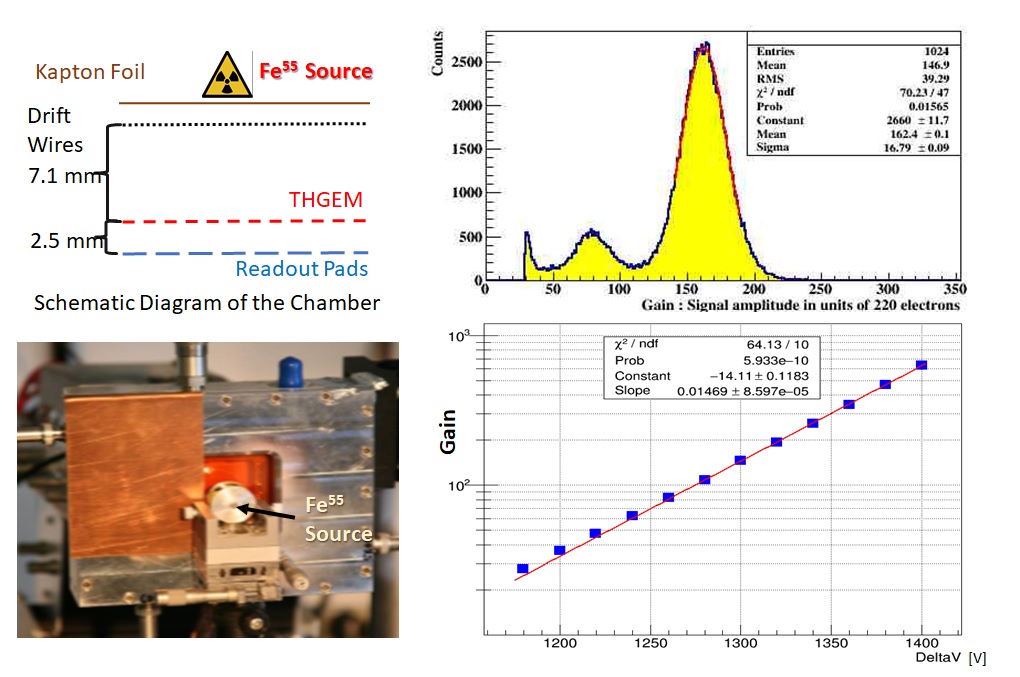}
\end{minipage}\hfill
\begin{minipage}[c]{0.98\textwidth}
        \caption{Top-left panel: The schematic of our detector assembly. Bottom-left panel: The detector is illuminated with an ${}^{55}Fe$ X-ray source . Top-right panel: A typical ${}^{55}Fe$ X-ray spectrum obtained in  $Ar-CO_{2}~(70\%-30\%)$ gas mixture when the  applied  voltages at drift, top and bottom of THGEM are -2520 V, -1720 V and -500 V respectively, while the anode is at ground. Bottom-right panel: shows the gain dependence of the THGEM versus the applied voltage.}
        \label{fig:Schematic_of_Detector_setup}
    \end{minipage}
\end{figure}

Each  THGEM is characterized in the setup schematized in figure  ~\ref{fig:Schematic_of_Detector_setup}. 
A plane of drift wires above it and a segmented readout anode plane, both properly biased, provide the drift and induction field respectively. The detector is operated with various gas mixtures, all including Ar. The electrons from $^{55}Fe$ converted by Ar are collected and multiplied in the hole region of the THGEM. The electron avalanche generated in the multiplication process, while drifting towards the anode,  induces 
the detected signal. 

All THGEMs used for our studies have been characterized using different gas mixtures at INFN Trieste before applying coating procedures: the goal is to perform comparative studies after coating  them with UV sensitive films.  
\par

 A typical ${}^{55}Fe$ X-ray spectrum obtained in  $Ar-CO_{2}~(70\%-30\%)$ gas  mixture is shown in   figure~\ref{fig:Schematic_of_Detector_setup}, top-right panel. The bottom right panel of figure \ref{fig:Schematic_of_Detector_setup} shows the gain dependence of THGEM versus the voltages applied between the two faces.

\subsection{Coating procedure}

\begin{figure}
\begin{minipage}[c]{1\textwidth}
    \includegraphics[width=\textwidth]{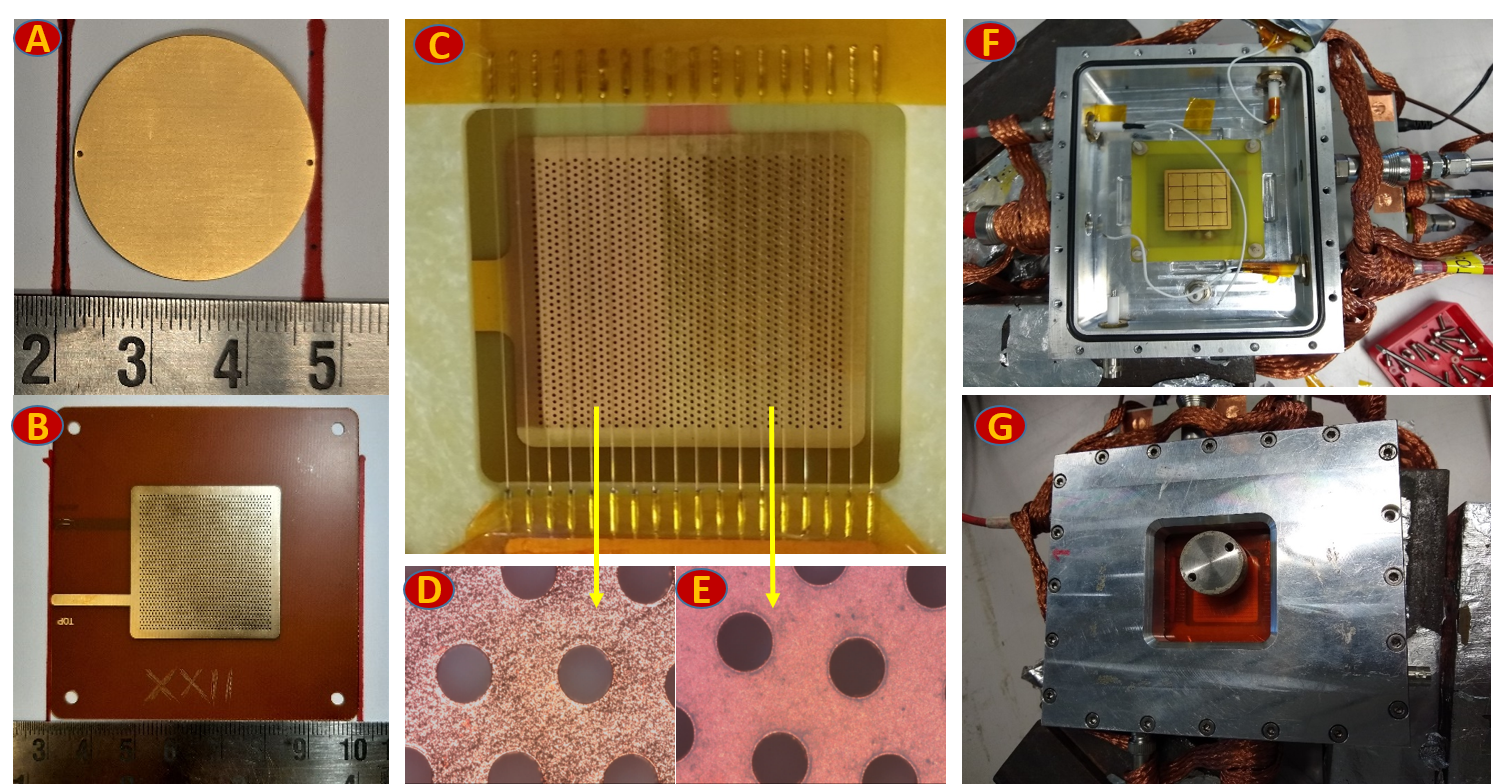}
\end{minipage}\hfill
\begin{minipage}[c]{0.98\textwidth}
        \caption{ (A) Au\_PCB of 1 inch diameter substrate used for the QE measurement. (B)  Uncoated THGEM of active area 30~mm$\times$30~mm. (C) Half uncoated and half coated THGEM, mounted into the test chamber and zoomed view of the both coated (D) and uncoated (E) part. (F) test chamber with readout pad where the THGEMs are tested. (G) The test chamber after installation of a THGEM, illuminated by an  ${}^{55}Fe$ X-ray source.}
        \label{fig:Detector_Images}
    \end{minipage}
\end{figure}

THGEMs have been coated either with raw ND, namely as-received powder, or with H-ND.  ND powder with an average grain size of 250 nm produced  by Diamonds \& Tools srl has been employed. The coating is covering either the whole surface of one of the THGEM faces, or half of it.
\par
The standard procedure of hydrogenation of ND powder photocathodes is performed by using the MicroWave Plasma Enhanced Chemical Vapor Deposition (MWPECVD) technique. For the treatment in microwave (mw) $H_{2}$ plasma, 30 mg of ND powder was placed in a tungsten boat (overall length 32 mm, trough 12 mm long × 5 mm wide × 1 mm deep, Agar Scientific Ltd) positioned on a heatable substrate holder of an ASTeX-type reactor evacuated to a base pressure better than $7\times{10}^{-7}$ mbar. The powder was heated to $650~^{0}$C using an external radiative heater (via a Proportional-Integral-Derivative feedback control system), then $H_{2}$ gas was flowed in the chamber at 200 sccm, the pressure and the mw power were maintained at 50 mbar and 1250 W, respectively. The heating due to the mw power increases further the temperature of the powders up to $1138~^{0}$C as determined by a dual wavelength ($\lambda _{1}$ = 2.1 $\mu m$ and $\lambda _{2}$ =2.4 $\mu m$) infrared pyrometer (Williamson Pro 9240). After 1 h of $H_{2}$ plasma exposure, the hydrogenated powder were cooled to room temperature under high vacuum.This procedure can not be used for THGEMs which are made by fiberglass, which does not tolerate temperatures above $180~^{0}$C.

This limitation is overcome by the novel and low-cost technique developed at INFN Bari  ~\cite{coating-I, coating-II}. The H-ND is obtained by treating the as-received powder in $H_{2}$ microwave plasma for one hour before deposition.

The ND and H-ND powder were separately dispersed in deionized water and sonicated for 30 minutes by a Bandelin Sonoplus HD2070 system. Then, the emulsion was sprayed on the THGEM at $120~ ^{0}$C or slightly higher temperature.  
\par 
Four THGEMs with different geometrical characteristics have been coated as listed below:
\begin{itemize}
\item ~0 $\mu m$ rim  - ND half coated
\item ~0 $\mu m$ rim  - H-ND half coated
\item 10 $\mu m$ rim  - H-ND full coated 
\item20 $\mu m$ rim   - ND half coated 
\end{itemize}
 A fifth THGEM with 10 $\mu m$ rim  was coated with a reflective CsI film by thermal evaporation technique at INFN  Bari. 
 \par
 Images of the coated substrates and the setup for the characterization are provided in Fig.~{\ref{fig:Detector_Images}}.


\begin{figure}
\includegraphics[width=.56\linewidth]{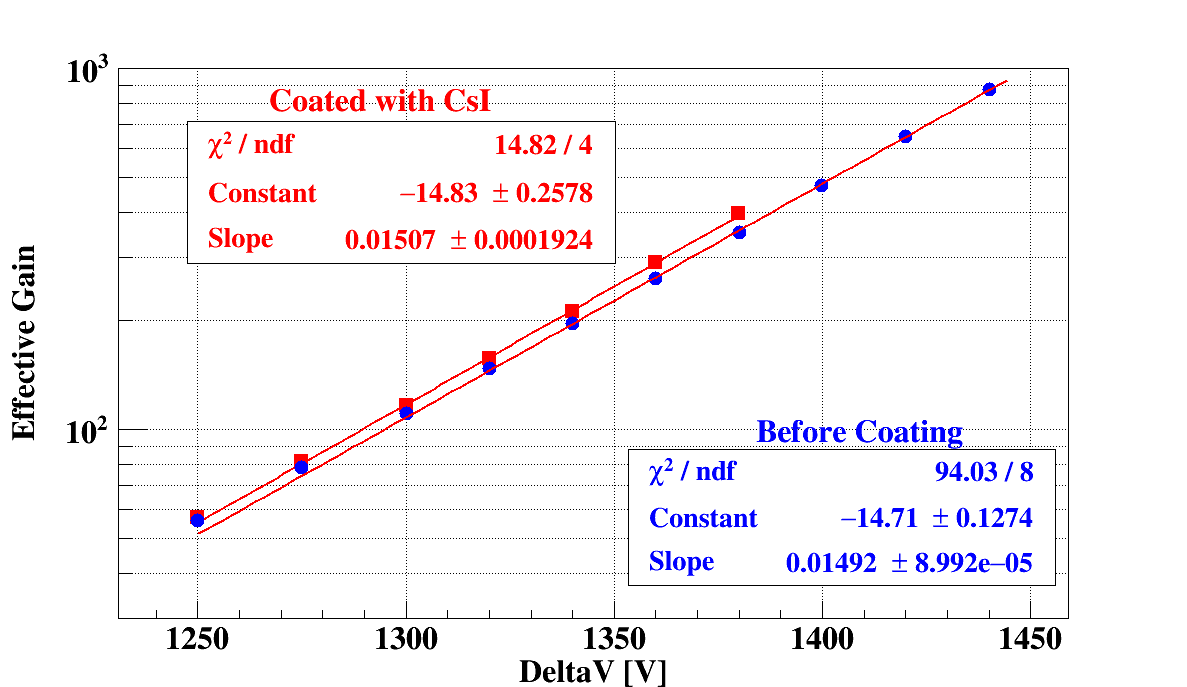}\hfill
\includegraphics[width=.44\linewidth]{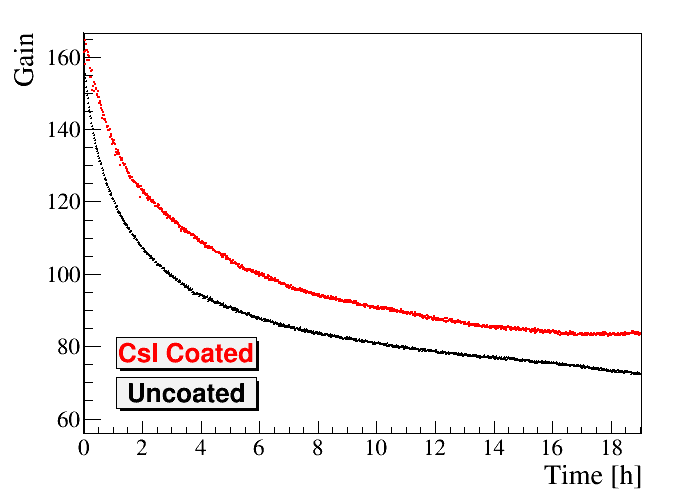}
\caption{The gain of the THGEM with 10 ~$\mu$m rim measured before and after CsI coating are compared. Left panel: Gain versus applied voltage across the THGEM electrodes. Right panel: gain evolution versus time.}
\label{fig:CsI_Comp}
\end{figure}

\subsection{Post-coating THGEM characterization}

The characterization of THGEMs coated with CsI, ND and H-ND provides interesting indications. The gain in the coated part tends to be  larger than the gain for the uncoated part in all the three cases. However, the increase for the coated part depends on the coating materials as well as on rim size.
\par
 
 The THGEM with $\sim10~\mu$m rim size, coated with a reflective CsI showed a 20\% gain increment in comparison to the uncoated ones as shown in Fig. ~\ref{fig:CsI_Comp}.
A tentative explanation of the observed gain increase is the lower rate of charging-up of the free dielectric surface. The surface resistivity is decreased when  the coating is present due to the resistivity of the coating film. 
\par
\begin{figure}
\includegraphics[width=.485\linewidth]{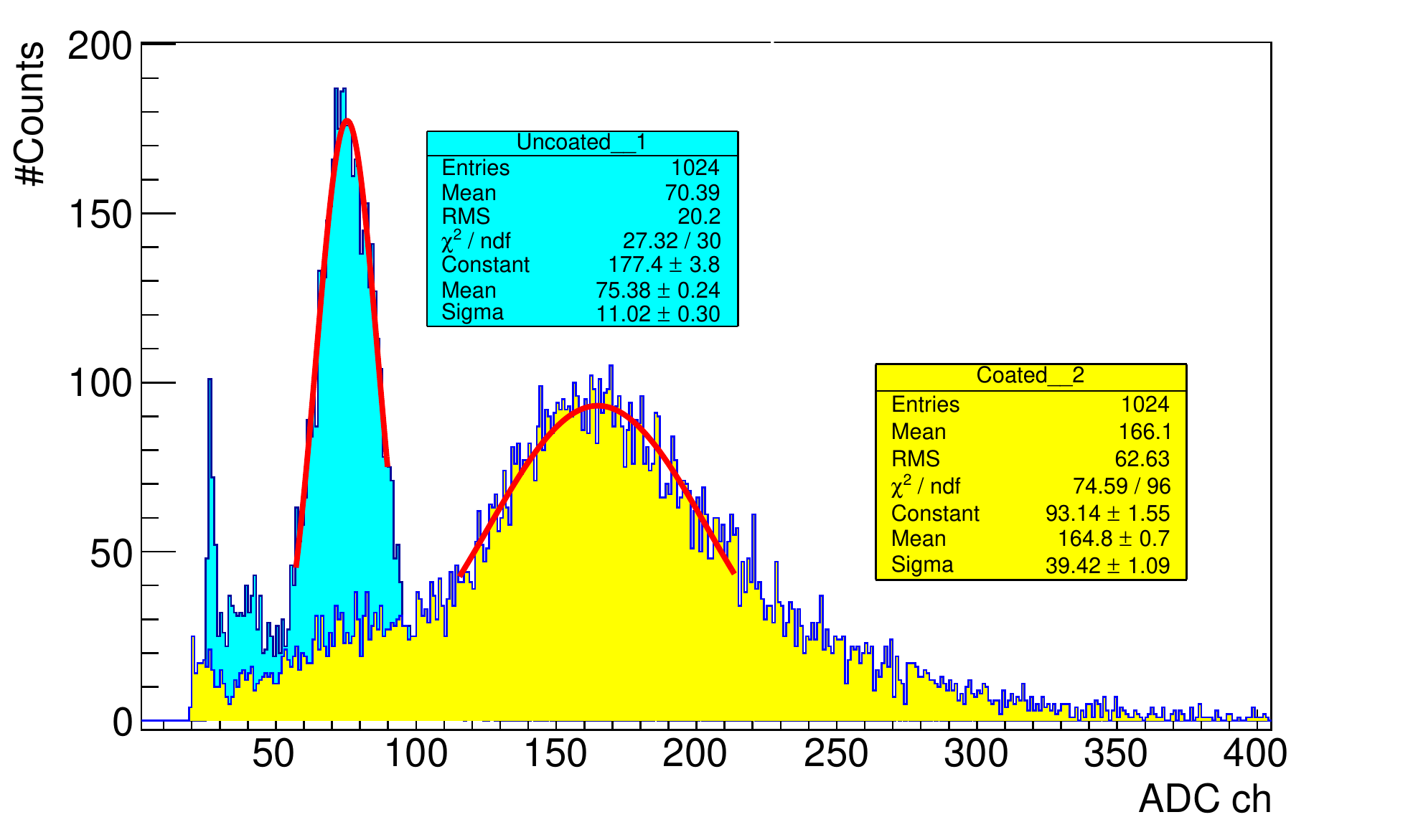}
\includegraphics[width=.48\linewidth]{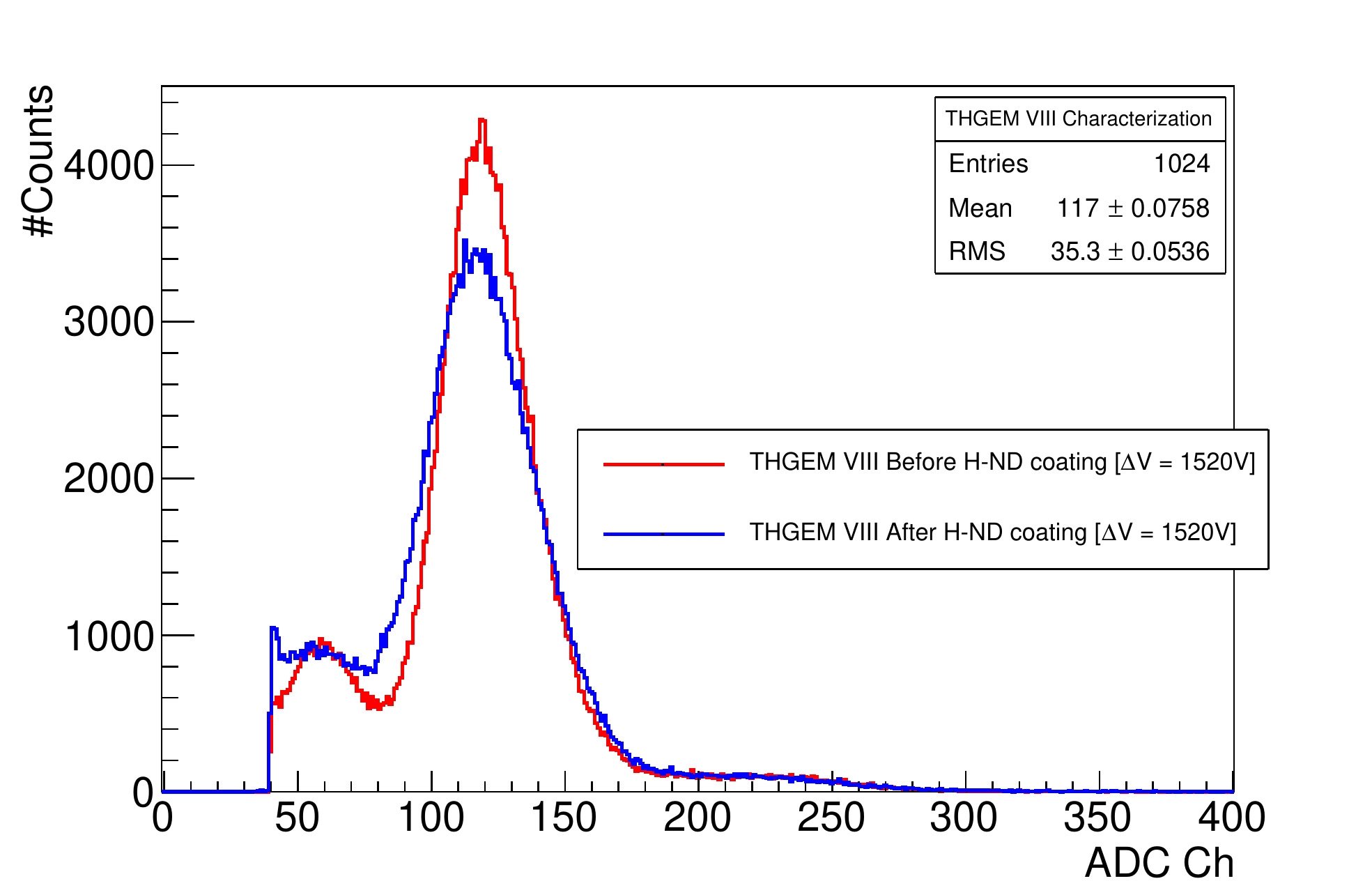}
\caption { Left panel: ${}^{55}Fe$ X-ray spectra obtained with a 20~$\mu$m rim THGEM half-coated with ND powder in  $Ar-CO_{2} (70\%-30\%)$ gas  mixture.  The voltages applied  to drift, top and bottom of the THGEM electrodes are 3510 V, 2110 V and 750 V respectively, while the anode is kept at ground. Right  panel: ${}^{55}Fe$ X-ray spectra obtained with a 0~$\mu$m rim uncoated THGEM. The same measurement after  coating the same THGEM with a H-ND emulsion prepared 17 months earlier in a $Ar-CH_{4}~(50\%-50\%)$ gas  mixture}
\label{fig:X_Ray_Spectra}
\end{figure}

The left panel of Fig. ~\ref{fig:X_Ray_Spectra}, shows the amplitude distribution for both uncoated and ND coated parts of the THGEM with $\sim20~\mu$m  rim size. The voltages applied  to drift, top and bottom of the THGEM electrodes are 3510 V, 2110 V and 750 V respectively. The gain of the ND coated part is $\sim$ 2 times  higher compared to the one of the uncoated part.


\begin{figure}
\includegraphics[width=.48\linewidth]{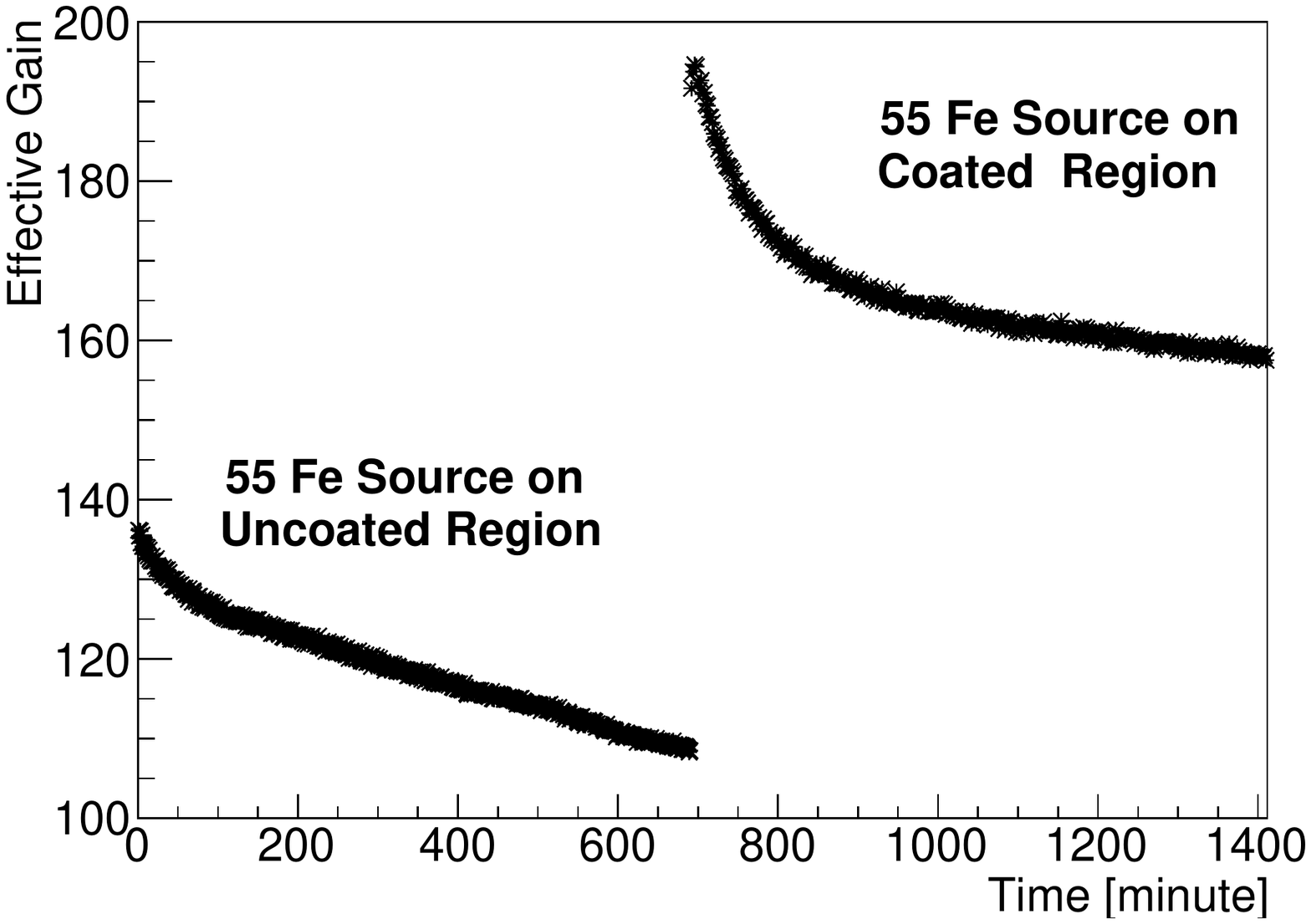}\hfill
\includegraphics[width=.48\linewidth]{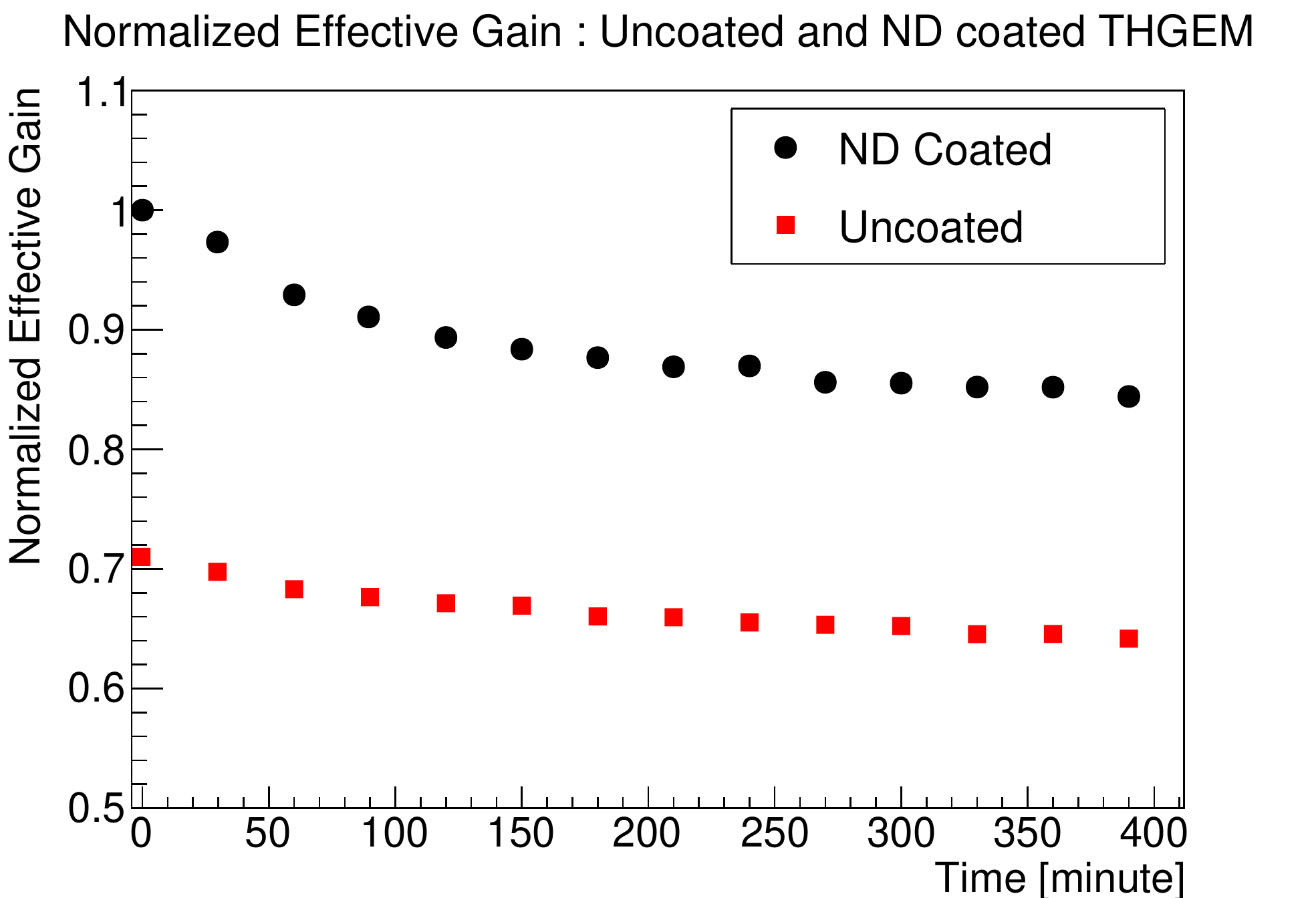}
\caption{Evolution versus time of the effective gain behavior of a THGEM with 0~$\mu$m rim, half-coated with ND. Gain versus time (left panel); the source is moved to the coated region at 700 minutes. The same data normalized to the maximum gain measured in the coating region versus time, where t=0 is when the illumination of a region starts (right panel).}
\label{fig:Nor_Gain_LR}
\end{figure}

In case of a ND coated THGEM with no rim the gain of the coated part is larger by a factor of $\sim$1.4 as shown in Fig.\ref{fig:Nor_Gain_LR} left panel. The gain is maximum when the X-ray source starts illuminating and it decreases gradually by ${\sim}$ 20\% in about 500 minutes. This effect is observed both for the ND coated and uncoated THGEM parts as illustrated in Fig.\ref{fig:Nor_Gain_LR}, right panel.
The tentative explanation of the gain increase is the same one already proposed for the gain increase observed with CsI coating. The increase is higher when the open dielectric surface is larger, namely for the $\sim20~\mu$m  rim THGEM.
\par


\par 
The H-ND coated THGEMs with 0 $\mu m$ and 10 $\mu m$ rim show a lower electrical stability as compared to the uncoated THGEMs and cannot be operated at nominal voltage. 
In order to study this unexpected behaviour a second exercise was performed using a new  THGEM with no rim fully coated with H-ND and $Ar-CH_{4}~(50\%-50\%)$ gas mixture. Immediately after the coating the THGEM could not be operated at the nominal voltage.  A heat treatment in an electric oven at ${120}~^{0}C$ for 24 hours allowed to perform the characterization. The right panel of figure ~\ref{fig:X_Ray_Spectra} shows the signal amplitude distributions measured before and after the H-ND coating followed by the heat treatment. No evidence for increase of gain is observed. This observation suggests that the electrical instability present before the heat treatment can be related to water molecules present at the H-ND surface.
\par
A consistent picture is emerging in spite of the initially unexpected results obtained characterising the coated THGEMs. The consolidation of this picture requires
further investigation.

\subsection{Ageing test of H-ND PC}
QE measurements can be performed with the available setup only in case of small-size samples. Therefore, the same coating procedure used for the THGEM samples has been applied to disk substrates, one inch diameter. The disk material and surface preparation are the same  of the THGEMs, even if no hole structure is present (Fig.~\ref{fig:Detector_Images} A). The H-ND coating is with an emulsion of  ND powder hydrogenated 17 months  earlier.

\begin{figure}
\begin{minipage}[c]{1\textwidth}
    \includegraphics[width=\textwidth]{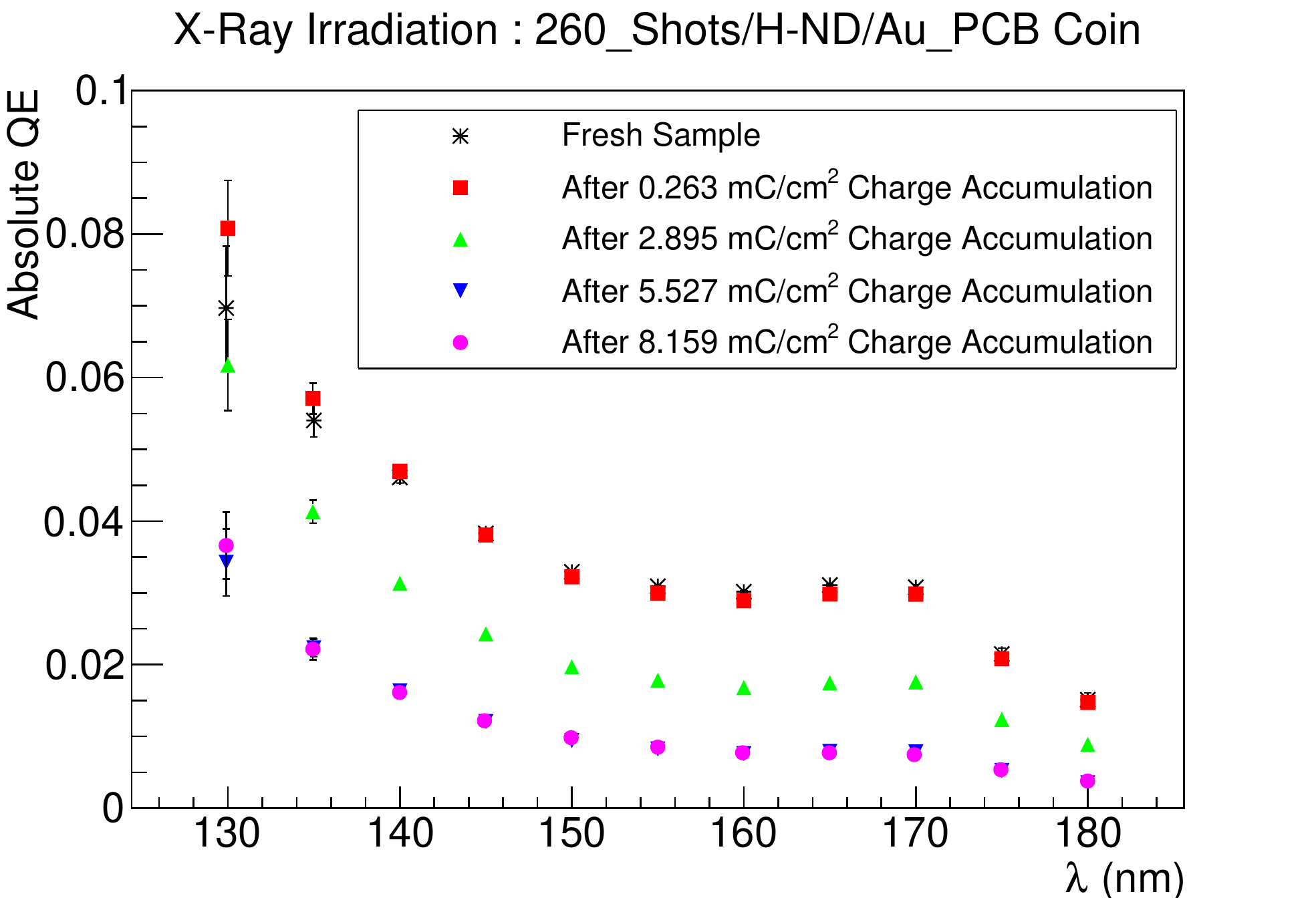}
\end{minipage}\hfill
\begin{minipage}[c]{0.98\textwidth}
        \caption{Quantum efficiency as a function of wavelength for fresh and various  charge  accumulations ($0.263mC{/}cm^{2}, ~2.895mC{/}cm^{2}, ~5.527mC{/}cm^{2} and ~8.159mC{/}cm^{2}$) due to ion bombardment on  H-ND coated Au\_PCB substrate.}
        \label{fig:Ageing_New}
    \end{minipage}
\end{figure}

The H-ND coated PCB disc was irradiated with a mini X-ray source in $Ar-CO_{2} (70\%-30\%)$ gas mixture. For the QE measurement a McPherson VUV monochromator (model 234/302) was employed. The RD-51 ASSET system~\cite{asset}  at CERN 
has been used. In this setup both the QE measurement and ion bombardment can be performed in the same system without exposure to air. Ions are generated by a gaseous multiplier metallic grid set at 5 mm from the disc surface and they impinge on the sample surface.
The QE of an  H-ND coated disc was measured before irradiation in a wavelength range from 130 nm to 180 nm with a scan step of 5 nm. The sample was then moved to the X-ray irradiation chamber using an automated manipulator under vacuum ($\approx {1\times{10}^{-7} mbar}$). The QE has been measured again after controlled doses of accumulated charge and the measurements are reported in Fig.~\ref{fig:Ageing_New}. 
The QE before irradiation and after an accumulated charge of $0.263~ mC{/}cm^{2}$ are  similar in whole wavelength range. This strongly supports the hypothesis that H-ND photocathode are more robust than CsI once respect to ion bombardment. In fact,  in case of a CsI photocathode, a 25\% drop in QE is observed for an irradiation of $1.0~ mC{/}cm^{2}$~\cite{NIMA_574_2007_28}. 
For the H-ND coated sample, a decrease in the QE of 42$\%$ and 74$\%$ was observed as charge accumulation reached the values of  $2.895~ mC{/}cm^{2}$ and $5.527 ~mC{/}cm^{2}$, respectively. Interestingly, we did not observe any further degradation in the QE for an  accumulated charge of $8.159~ mC{/}cm^{2}$.
This is the first  preliminary irradiation ageing study of H-ND photocathodes ever performed.


\section{Conclusion}

THGEM samples  coated with different types of photosensitive layers (CsI, ND and H-ND) have been studied. An increase in the gain response for the CsI and ND coated THGEMs was observed compared to the uncoated ones. The electrical instability of the H-ND coated THGEM, initially observed,  is overcome by a heat treatment. No gain enhancement is observed.  
\par
The X-ray irradiation study on H-ND photocathodes performed by us for the first time indicates that H-ND is more robust than CsI. We can conclude that H-ND photocathode material is a promising alternative to CsI based photocathodes for all applications in the far VUV domain requiring high robustness.

\section{Acknowledgment}
This R\&D activity is partially supported by
\begin{itemize}
\item 
EU Horizon 2020 research and innovation programme, STRONG-2020 project, under grant agreement No 824093;
\item 
the Program Detector Generic R\&D for an Electron Ion Collider by Brookhaven National Laboratory, in association with Jefferson Lab and the DOE Office of Nuclear Physics.
\end{itemize}

\end{document}